\documentclass[11pt,preprint]{aastex}
\usepackage{chngpage}
\usepackage{graphicx}
\usepackage{natbib}

\shorttitle{Delayed Energy Injection Model}
\shortauthors{Geng \& Wu}

\begin{document}

\title{Delayed Energy Injection Model For Gamma-Ray Burst Afterglows}

\author{J. J. Geng\altaffilmark{1, 2}, X. F. Wu\altaffilmark{3, 4, 5}, Y. F. Huang\altaffilmark{1, 2} and Y. B. Yu\altaffilmark{1, 2}}

\altaffiltext{1}{Department of Astronomy, Nanjing University, Nanjing 210093, China; hyf@nju.edu.cn}
\altaffiltext{2}{Key Laboratory of Modern Astronomy and Astrophysics (Nanjing University), Ministry of Education, China}
\altaffiltext{3}{Purple Mountain Observatory, Chinese Academy of Sciences, Nanjing 210008, China; xfwu@pmo.ac.cn}
\altaffiltext{4}{Chinese Center for Antarctic Astronomy, Chinese Academy of Sciences, Nanjing 210008, China}
\altaffiltext{5}{Joint Center for Particle Nuclear Physics and Cosmology of Purple Mountain Observatory¨CNanjing University, Chinese Academy of Sciences, Nanjing 210008, China}

\begin{abstract}
The shallow decay phase and flares in the afterglows of gamma-ray bursts (GRBs) is widely believed to be associated with the later activation of central engine. Some models of energy injection involve with a continuous energy flow since the GRB trigger time, such as the magnetic dipole radiation from a magnetar. However, in the scenario involving with a black hole accretion system, the energy flow from the fall-back accretion may be delayed for a fall-back time $\sim t_{\rm fb}$. Thus we propose a delayed energy injection model, the delayed energy would cause a notable rise to the Lorentz factor of the external shock, which will ``generate'' a bump in the multiple band afterglows. If the delayed time is very short, our model degenerates to the previous models. Our model can well explain the significant re-brightening in the optical and infrared light curves of GRB 081029 and GRB 100621A. A considerable fall-back mass is needed to provide the later energy, this indicates GRBs accompanied with fall-back material may be associated with a low energy supernova so that fraction of the envelope can be survived during eruption. The fall-back time can give meaningful information of the properties of GRB progenitor stars.
\end{abstract}


\section{Introduction}
\label{sect:intro}

The multi-wavelength observation of gamma-ray bursts (GRBs) by {\it Swift}~satellite has opened a new era for GRB studies. Many unexpected phenomenons in the X-ray and optical light curves have been collected. In summarizing the X-ray afterglow data, a canonical light curve in the X-ray band may consist five components (Zhang et al. 2006), of which two segments intrigue many theorists. The first is the shallow decay phase (we can refer to it as a plateau), usually lasts from $10^{3}$ to $10^{4}$ s, during which the flux declines slowly with time. The second is the sudden rise and precipitous drop features `superposed' on the power-law X-ray light curve, which are referred to as flares. Both these two segments deviate from a smooth power-law evolution predicted in the frame of standard external shock afterglow model (Piran et al. 1993; Sari \& Piran 1999a,b). The late-time re-brightenings in some GRB optical afterglows also show this deviation (Nardini et al. 2011a,b; Greiner et al. 2013). In the frame of external shock fireball model, the optical re-brightening may associate with the discontinuity in the external medium density profile (e.g., Dai \& Wu 2003; Lazzati et al. 2002; Nakar \& Piran 2003) or possible variations of the micro-physical parameters of the fireball (Kong et al. 2010). However, numerical simulations have shown that even a sharp discontinuity with a strong increase in the density profile cannot produce a sharp bump in the light curve and only a smooth and diluted change can be observed (Nakar \& Granot 2007).

The X-ray plateau and X-ray flares strongly indicate that the inner engines of GRBs should be active much longer than the prompt gamma-ray emission (Zhang 2007; Lazzati \& Perna 2007), or the energy release can be restarted from the central engine for some mechanism (e.g., Proga \& Zhang 2006). As many models of GRBs involve accretion onto a compact object, usually a black hole (Woosley 1993; MacFadyen \& Woosley 1999; Popham et al. 1999; Narayan et al. 2001; Tutukov \& Fedorova 2007). The late activities of the black hole (BH) may be sustained by the accretion of the fall-back material which fails to escape from the progenitor star (Perna et al. 2006; Kumar et al. 2008a). In the proto-magnetar model (Metzger et al. 2011), a rapidly spinning, strongly magnetized proto-neutron star produces outflows, which give birth to the prompt emission. While the magnetar cools, the wind becomes ultra-relativistic and Poynting-flux dominated, this may continue to power the late time flaring or afterglow emission. If the assumption of later energy injection is true, the interaction between the injected flow and the external shock may explain the mysterious features of X-ray afterglows. For example, the pure electromagnetic injection model has been initially proposed to interpret the X-ray plateau (Dai \& Lu 1998a,b; Fan \& Xu 2006; Kong \& Huang 2010; Xue et al. 2009). Zhang et al. (2006) considered the continuous energy injection into the fireball during the deceleration phase, and gave an analytical result of the temporal index if the central engine is long lasting, behaving as $L (t) \propto t^{-q}$. A more physical version of the energy injection scenario of a pulsar assumes that the pulsar may continuously eject an ultrarelativistic electron-positron-pair wind, the wind will interact with the external medium, during which the reverse shock emission can account for plateaus in some GRBs (Dai 2004; Yu \& Dai 2007; Mao et al. 2010). The later X-ray flares can also be explained by later internal shock model (Wang \& Cheng 2012; Zou et al. 2013), in which the long shock catching time ``defers'' the X-ray emission.

However, if the energy injection is somehow delayed (the activity of the central engine is suspended after the prompt emission and restarts again later; e.g., Proga \& Zhang 2006; Lei et al. 2008), the shock dynamics may evolve from a non-injection phase to the injection-dominated phase. During this transition, a rapid change on the Lorentz factor $\gamma$ of the external shock will occur on a time scale $\delta t < t$. Steep rise of $\gamma$ will naturally lead to a flux rise, corresponding to the flares in some X-ray or optical light curves. Thus we propose a delayed energy injection model to interpret the bumps in X-ray or optical band. In this model, we assume there exits a start time $t_{\rm s}$ (in the observer frame) from which the energy flow begins to inject into the external shock produced before. The dynamics of the outflow is described by the equations in Huang et al. (1999, 2000a,b) and the revised version by Kong \& Huang (2010) and Xue et al. (2009). We take the form of the energy flow as poynting-flux, this is applicable since the numerical results from the poynting-flux injection or the pair wind injection differ little with each other in the soft X-ray band (Yu \& Dai 2007). Soon after the injection, the shock dynamics will approach the trace predicted by the normal injection models proposed before. So our model can also give an explanation to the plateau phase following the bump.

In our study, we select GRB 081029 and GRB 100621A as examples of which the multi-wavelength observation data are of high quality. GRB 081029 and GRB 100621A are both characterized by the fast re-brightening in the optical afterglow (Nardini et al. 2011a,b; Greiner et al. 2013). Within our model, the afterglow can be well fitted. Our paper is organized as follows. In Section 2, we describe our delayed energy injection model and some numerical results are showed. In Section 3, we describe the observations of these two GRBs and apply our model to interpret them. Our conclusions are summarized in Section 4.

\section{Delayed energy injection model}
\label{sect:model}

\subsection{Shock dynamics}

A generic dynamical model of GRB outflow was proposed by Huang et al. (1999, 2000a), and has been widely used to interpret the afterglow light curves (e.g., He et al. 2009; Xu \& Huang 2010). Recently, Pe'er (2012) and Nava et al. (2013) have corrected the equation by considering the role of the adiabatic and radiative losses on the hydrodynamical evolution of the shock. However, the numerical results are little different from that of Huang et al. (1999, 2000a). Here for simplicity, we still use the old equation. When Poynting-flux energy injection is taken into account, the basic equation for GRB outflow dynamics during the afterglow phase proposed by Huang et al. (2000a) can be modified to be (also see Kong \& Huang 2010; Xue et al. 2009)
\begin{equation}
\frac{d \gamma}{d m} = - \frac{(\gamma^{2} - 1)-\frac{1 - \beta}{\beta c^{3}} \Omega_{j} L(t - R/c) \frac{d R}{d m}}{M_{ej} + 2 (1 - \varepsilon) \gamma m + \varepsilon m},
\end{equation}
where $\gamma = 1 / \sqrt{1 - \beta^{2}}$ is the bulk Lorentz factor of the shocked medium, $\Omega_{j} = (1 - \cos \theta_{j}) / 2$ is the beaming factor of the GRB outflow, $\theta_{j}$ is the half-opening angle of the jet, $M_{ej}$ is the initial mass of the jet, $m$ is the swept-up mass by the shock, $\varepsilon$ is the radiative efficiency, $R$ is the radius. The forward shock may be continuously refreshed with the additional energy $L$, thus the emission from the electrons accelerated by the shock can decrease very slowly or even rise in some cases.

\subsection{External shock radiation}

We briefly describe the radiation process from the external forward shock based on the standard model (Sari et al. 1998) in this subsection. Here, we will use prime (') on variables to denote parameters in the shock comoving frame and characters without a prime to denote parameters in the observer frame. The distribution function $d n_e'/ d \gamma_e'$ of the injected electrons is often taken as a power-law form with the index $-p$. Then the minimum Lorentz factor $\gamma_m'$ of shock accelerated electrons is
\begin{equation}
  \gamma_i' = {(p-2)m_p\over (p-1)m_e} \epsilon_e (\gamma -1),
\end{equation}
and the cooling Lorentz factor $\gamma_c'$ is
\begin{equation}
\gamma_c' = {6 \pi m_e c\over \sigma_T B'^2 (1+Y) t'},
\end{equation}
where $ \epsilon_B$ and $\epsilon_e$ are shock equipartition parameters for magnetic fields and electrons, $B'$ is the magnetic field generated in the plasma, $Y$ is the energy ratio between the
inverse Compton component and the synchrotron component.

In the standard model, the main radiation mechanism for the outflow is the synchrotron radiation from electrons (Sari et al. 1998; Sari \& Piran 1999a,b). In this context, the synchrotron emission from electrons can be approximated by a broken power-law spectrum with two characteristic break frequencies corresponding to the characteristic electron Lorentz factor: $\nu_m = 3/2 \gamma \gamma_m'^2 \nu_L'$ and $\nu_c = 3/2 \gamma \gamma_c'^2 \nu_L'$, where $\nu_L'$ is the Larmor frequency. The evolution of $\nu_{m}$ and $\nu_{c}$ determines the evolution of the radiation spectrum, and then the light curve can be obtained. If the circumstance is homogeneous interstellar medium (ISM), the typical synchrotron frequency and the cooling frequency are $\nu_m \propto \epsilon_B^{1/2} \epsilon_e^2 t^{-3/2}$ and $\nu_{c} \propto (1+Y)^{-2} \epsilon_B^{-3/2} t^{-1/2}$ in the ``standard'' case (adiabatic evolution with prompt injection of energy). While in the wind type environment, $\nu_m \propto \epsilon_B^{1/2} \epsilon_e^2 t^{-3/2}$, $\nu_{c} \propto (1+Y)^{-2} \epsilon_B^{-3/2} t^{1/2}$. As we will see below, the delayed energy injection will modify the normal evolution of $\gamma$, so that $\nu_m$ and $\nu_c$ will experience a sharp transition between two phases. It is this transition that finally gives a consequence of X-ray and optical bump.

\subsection{Numerical result}

In previous work, a Poynting flux power luminosity $L = L_{0} (1 + t / T)^{-2}$ is often assumed to be continuously injected into the shock (Dai \& Lu 1998a; Zhang \& M{\'e}sz{\'a}ros 2001; Yu \& Dai 2007), where $L_{0}$ is the initial luminosity at $t = 0$, $T$ is the characteristic spin-down timescale of magnetar. However, this continuous energy can only give a consequence of mild deviation from canonical afterglow. For the violent bumps, a delayed injection may be needed. In the context of collapse of a massive star for long GRBs, the fall-back and accretion of the stellar envelop seems reasonable and can give additional consequences (Kumar et al. 2008a,b; Cannizzo et al. 2011; Wu et al. 2013). Since we know little information of the process of fall-back accretion, the injected luminosity may be of various forms. Here, we focus on two possible modes: ``top-hat'' mode and broken-power-law mode, detailed description of these modes and the corresponding consequences are given below.

\subsubsection{``top-hat'' mode}

This mode refers to a constant injection mode showed in the left panel of Figure~1. The injected power retains to be a constant from $t_{\rm s}$ to $t_{\rm e}$ (in the observer frame). The simulation result from Kumar (2008b) shows a plateau of luminosity produced by continued fall-back of matter, so this mode seems applicable to the cases with a plateau phase in the afterglow. By taking $L = L_{0}$ (defined at the cosmological local frame) in Equation (1), we can calculate the evolution of the Lorentz factor in our model, then afterglows. In our calculations, the typical values (e.g., Huang et al. 2000b; Freedman \& Waxman 2001; Wu et al. 2003) adopted for parameters of the afterglow model are $E_{K,{\rm iso}} = 1.0 \times 10^{52}$ erg, $\theta_j = 0.1$ rad, $p = 2.3$, $\epsilon_e = 0.1$, $\epsilon_B = 0.01$, $\Gamma_0 = 300$, $n = 1.0$ cm$^{-3}$ for the ISM case, where $E_{K,{\rm iso}}$ is the isotropic explosion kinetic energy, $\Gamma_0$ is the initial Lorentz factor of jet and $n$ is the density of the circumburst environment. If the environment is wind type, the density is characterized by the parameter $A_* = 0.5$ ($n = 3 \times 10^35 A_* / r^2$ cm$^{-3}$) in our calculation. For the injection process, we take $L_0 = 1.0 \times 10^{50}$ erg~s$^{-1}$ (istropic), $t_{\rm s} = 5000$ s, $t_{\rm e} = 6000$ s.

Figure~2 shows the numerical results by taking the parameters above, the left panel and right panel correspond to the ISM case and wind case, and the top panel and lower panel correspond to the evolution of characteristic frequencies and afterglows in two observable bands ($4.0 \times 10^{14}$ Hz as optical band and $0.3$ keV as soft X-ray band). In the top left panel of Fig. 2, $\nu_m \propto t^{-3/2}$ and $\nu_c \propto t^{-1/2}$ before the energy injection, then $\nu_m$ and $\nu_c$ experience a quick evolution when the later energy begins to change the shock dynamics. They then slowly transit to a new trajectory like the canonical one before. Therefore there is a period for these two characteristic frequencies to transit from the normal evolving phase to the phase dominated by injected energy. This period corresponds to later re-brightening of afterglows showed in the lower left panel. Besides, in this case, $\nu_c$ goes across the optical band and gets higher again during the injection period, the optical spectral index should become softer due to the injection and get harder back soon. This can interpret the possible spectrum evolution within the bump. The right panel in Fig. 2 makes the same sense for wind case.

In our calculations, the effect of equal time arrival surface (ETAS; e.g., Waxman 1997; Granot et al. 1999) has been considered, so the light curve seems to be relatively smooth rather than very sharp. Another consequence brought by ETAS is that the peak time of the high energy band and the the low energy band is different (Huang et al. 2007). This will be also showed in Section 3. One should notice that in the wind case, if the parameters are ``selected'' appropriately so that $\nu_m$ and $\nu_c$ crosses each other shortly after the bump, there will be a short plateau or even small bump following the ``big'' bump before. This is because the cooling Lorentz factor is calculated with different equations before and after the crossing time due to the synchrotron self-Compton (SSC, e.g., Sari \& Esin 2001) process. $\nu_c$ will have a small change after the crossing time. This result can be potentially used to interpret the small structure after the bump. It is also noticeable that a relative lower kinetic energy $E_{K,{\rm iso}}$ is essential to ensure the notable change of $\gamma$ during the injection of a limited luminosity. This indicates that the effect of energy injection would be significant only for low luminous GRBs, otherwise the fall-back energy should be large enough.

In principle, the spectrum of the afterglows can be modified markedly by $t_s$ in our delayed energy injection model, by affecting the evolution of $\nu_m$ and $\nu_c$ before (using the fireball parameters) and after (using the injection parameters) $t_s$. So these parameters can be constrained from the spectrum evolution around $t_s$ if detail observations are available.

\subsubsection{broken-power-law mode}

This mode refers to a more realistic case in which the luminosity rises and declines with time. We assume the fall-back accretion starts at a fall-back time $t_{\rm s} = t_{\rm fb}$ (defined at the cosmological observer frame). According to the previous work (MacFadyen et al. 2001; Zhang et al. 2008; Dai \& Liu 2012; Wu et al. 2013), during fall-back accretion process, the fall-back accretion rate $\dot{M}$ evolves with time as $\dot{M} \propto t^{1/2}$ before the peak time $t_{\rm p}$, and $\dot{M} \propto t^{-5/3}$ in the late time. To convert the $\dot{M}$ to the output power $L$, one need to know details about the mechanism of relativistic jet launch. For the BH accretion system as central engine of GRBs, the jet power are mainly from two ways: the magnetic processes through the Blandford-Znajek (Blandford \& Znajek 1977; Lee et al. 2000) mechanism, Blandford-Payne (Blandford \& Payne 1982) mechanism or the disk magnetic reconnection (Yuan \& Zhang 2012); and the annihilation of neutrinos and anti-neutrinos in the neutrino-dominated accretion flows (Narayan et al. 2001). Many authors have attempted to investigate these processes (e.g., Popham et al. 1999; Xie et al. 2009; Lei et al. 2009, 2013), however, the jet efficiency $\eta = L / \dot{M} c^2$ is still mysterious up to now (McKinney 2005; Narayan et al. 2013).

Here, for simplicity, we assume the late energy release is proportional to $\dot{M}$, then we can obtain the luminosity profile of the fall-back power,
\begin{equation}
L = L_{\rm p} [\frac{1}{2} (\frac{t - t_{\rm s}}{t_{\rm p} - t_{\rm s}})^{- \alpha_r s} + \frac{1}{2} (\frac{t - t_{\rm s}}{t_{\rm p} - t_{\rm s}})^{- \alpha_d s}]^{-1/s},
\end{equation}
where $L_{\rm p}$ is the peak luminosity at the peak time $t_{\rm p}$, $\alpha_{r}$, $\alpha_{d}$ are the rising and decreasing index respectively, $s$ is the sharpness of the peak. This kind of profile is showed in the right panel of Fig. 1. For the realistic case, the injected power should depend on some more physical variables (like BH spin parameter $a$, the magnetic field $B$ around the BH). However, the realistic luminosity profile should essentially include a rising and a decreasing segment, so our simplification can roughly describe the jet luminosity during the fall-back accretion.

In our calculations, the typical values of fireball model parameters are the same as those in the ``top-hat'' mode. While for the injection process, we take $L_{\rm p} = 1.0 \times 10^{50}$ erg~s$^{-1}$, $t_{\rm s} = 5000$ s, $t_{\rm p} = 5200$ s, $t_{\rm e} = 5600$ s, $\alpha_r = 0.2$, $\alpha_d = -1.2$ and $s = 0.5$. Figure~3 shows the numerical results in the broken-power-law mode. It can be seen that the result of this mode differs little from that in the ``top-hat'' mode, this is mainly because both these two modes lead to the same evolution of $\gamma$.

\section{Application to GRBs}
\label{sect:appl}

\subsection{GRB 081029}

GRB 081029 was detected by the {\it Swift} satellite at 01:43:56 UT on 29 October 2008, it is a long-soft burst with a redshift of $z = 3.8479$ (D'Elia et al. 2008). The X-ray Telescope (XRT; Burrows et al. 2005) started to observe the field of GRB 081029 2.7 ks after the BAT trigger. Figure~4 shows the overall temporal evolution of afterglow in X-ray and optical band. The X-ray light curve can be modeled by a broken power law ($f(t) \propto t^{- \alpha}$). The best fit has decay indices are $\alpha_{X,1} = 0.56 \pm 0.03$ and $\alpha_{X,2} = 2.56 \pm 0.09$, with a break time of $t_{X,b} = 18230 \pm 346$ s (Holland et al. 2012). The initial X-ray light curve shows evidence of possible flares between 2550 and 5000 s after the trigger time. However, the lack of X-ray data during this period missed the possible flares. The X-ray spectrum can be fitted by a power-law with $\beta_{X} = 0.98 \pm 0.08$ ($f_{\nu} \propto \nu^{\beta}$). The optical and infrared data is from the observations of the Gamma-Ray burst Optical and Near-infrared Detector (GROND). Between about 3000 and 5000 s, the optical and infrared light curves rise rapidly (clearly seen in the GROND data; Nardini et al. 2011a), the rising index is $\alpha \simeq -8$ during this period. The rise of the optical re-brightening is so steep that one can nearly rule out the explanations of the discontinuity in the external medium density profile and the two component jet model (Filgas et al. 2012). A later reactivation of the central engine are preferred (Nardini et al. 2011a). Yu \& Huang (2013) have also given a numerical fitting to the afterglow of  GRB 081029 using a two-step injection model. However, they have only fitted the light curves in two bands and the SSC process has not be considered in their calculation.

Since our model can produce giant bumps in the optical bands as showed in Fig. 2 and Fig. 3, we now apply the model to fit the multi-wavelength afterglow of GRB 081029. The shallow decay and small bump after the peak of the bump strongly indicate that the circumstance should be wind type according to the comparison in Figures~2 and 3. Although we can ``construct'' one case in which $\nu_m$ crosses $\nu_c$ just after the peak time and one small bump (at $~ 8000$ s in the afterglows) would emerge naturally in the wind case, there is a technical problem for us to adjust this small numerical bump to the data to get a good fitting. At last we attribute this small feature to some other physical processes (e.g., the variability of the injected power due to some instabilities in the central engine) and use twice injection processes in the ISM case to interpret the observation.

The fitting results are showed in Fig. 4. We do not include the data in $g'$ and $r'$ bands since they may be strongly affected by the Lyman break at z=3.8 (Nardini et al. 2011a). The first injection is assumed to be in broken-power-law mode starting from $3300$ s, and the second injection is assumed to be the ``top-hat'' mode starting from $10000$ s. Other parameters of the twice injection are summarized in Table~1. The reduced chi-squared $\chi^2 / \nu$ of the X-ray fitting in the low panel of Fig. 4 is 1.87. The $\chi^2 / \nu$ of fitting for other five bands are all larger than 10 due to the small structures of the light curves and the small error bars of the observations. However, our main purpose is to reproduction the giant bump, and we realize it in this fitting. One consequence when taking effect of ETAS into consideration is that the peak times of bumps in different bands are actually slightly different (Huang et al. 2007), which can be seen clearly in the middle panel of Fig. 4. It can be easy understood that the refreshed shock with increasing $\gamma$ would prefer emitting high energy photons. This effect should also hold for the models involving refreshed shock based on external shock models.

\subsection{GRB 100621A}

GRB 100621A triggered the BAT on the {\it Swift} satellite on June 21, 2010 at 03:03:32 UT (Ukwatta et al. 2010). Its redshift is $z=0.542$ (Milvang-Jensen et al. 2010). The afterglow light curves of GRB 100621A were observed with XRT in X-rays and GROND in its seven filter bands (Greiner et al. 2013). This afterglow is more unusual than that of GRB 081029. The overall temporal evolution of afterglow at X-ray and the optical/NIR is showed in Figure~5. The optical/NIR light curves show a rapid rise with $\alpha \approx -4.0$ from 230 s after trigger. Since the early rise of on-axis canonical afterglow can only be $t^2$ or $t^3$ (Xue et al. 2009), this early rise is likely dominated by a flare (Greiner et al. 2013). Like GRB 081029, a steep rise during 4000-5000 s is also observed in the GROND band. Here, we would apply our delayed energy injection model to interpret this intensive jump. The data beyond 10 ks is higher than the extrapolation of the decay after jump, which indicates that there exists a component which dominate near 10 ks. We attribute this distortion to an off-axis afterglow emerging together with the on-axis one.

The curvature in the GROND spectrum illustrates the strong extinction of the afterglow light in the host galaxy (Figure~3 in Greiner et al. 2013). In order to fit the overall light curve of GRB 100621A, we should assume the extinction of the afterglow by the host galaxy. Also, we have assumed the spectrum of early flare as $f_{\nu} \propto \nu^{-1.3}$ and used the ``Norris function''{\footnote{Norris et al. (2005) has used function $f (t) = C \exp(-\tau_1/t-t/\tau_2)$ to fit pulse shape light curve, where $C$ represents the intensity, $\tau_1$ and $\tau_2$ are rise and decay timescales.}} (Norris et al. 2005) to describe the profile of the flare. Our fitting results are showed in Fig. 5. The injection of the on-axis jet is assumed to be in broken-power-law mode starting from 3600 s. Detailed parameters of the afterglow and the injection are summarized in Table~2, the dust extinction value of seven bands we used is listed in Table~3. The flux of the host galaxy is obtained using the later data, and it is added to the numerical results during fitting.

\section{Discussion and Conclusions}
\label{sect:disc}

The fireball model has well explained the ``clean'' broken-power-law afterglows before the {\it Swift} era. However, many unexpected structures like plateaus or bumps have been observed almost in half of the long GRBs (Gehrels et al. 2009). Models to explain the plateaus or bumps always involve in ``finding'' later energy sources which would dominate the later radiation. The Poynting-flux from magnetars (Metzger et al. 2011; Dai \& Liu 2012; Bernardini et al. 2012) or BH accretion systems (Cannizzo et al. 2011; Lindner et al. 2012) can play this role. When the later reactivities of the central engines have to be considered, the later radiation may be modified in two ways: refreshing the external shock in the fireball frame or attributing the later light curve to the jet luminosity directly from central engine. Here in this article, we consider the first case and give the numerical results based on two possible injection modes. Since we only consider the dynamic evolution of the forward shock changed by the later injection and other physical processes for photons streaming through the external shocks are neglected (Kumar et al. 2013), our model differs little from a ``second'' afterglow superposing on the original afterglow. However, this model can indeed explain the later bumps in the later multiple-wavelength observations of GRB 081029 and GRB 100621A as showed in Section 3.

The bumps in the later multiple-wavelength observations indicate the reactivity of the central engine. If this bump is truly due to the mass fall-back, the re-brightening time should correspond to the fall-back time $t'_{\rm fb} = t_s/(1 + z) \simeq (\pi^2 r^3_{\rm fb} / 8 G M_{BH})^{1/2}$, where $t'_{\rm fb}$ is defined in the cosmological local frame and $M_{BH}$ is the BH mass. For GRB 081029, $t_s = 3300$ s. This suggests the minimum radius around which the matter starts to fall back is $r_{\rm fb} \simeq 5.3 \times 10^{10} (M_{BH} / 3M_{\odot})^{1/3} (t'_{\rm fb} / 680 {\rm s})^{2/3}$ cm. In first principle, assuming the potential energy of the fall-back material is all converted to the jet power, we have $G M_{\rm BH} M_{\rm fb} /r_{fb} \simeq [L_{\rm p} (t_{\rm p} - t_{\rm s}) + L_0 (t_{\rm e} - t_{\rm s})] (1 - \cos\theta_j) / 2.0 / (1+z) $. The total fall-back mass is $M_{\rm fb} \simeq 3.5 M_{\odot} (M_{BH} / 3M_{\odot})^{-2/3}$, which seems large according to some articles (Lindner et al. 2010; Cannizzo et al. 2011). However, the mass should be overestimated since the energy of rotating BH and magnetic field around are also significant energy source. On the other hand, the large luminosity needed may indicate that our model is an inefficient scheme to use the injected energy. While for GRB 100621A, $r_{\rm fb} \simeq 1.2 \times 10^{11} (M_{BH} / 3M_{\odot})^{1/3} (t'_{\rm fb} / 2330 {\rm s})^{2/3}$ cm, the total fall-back mass is $M_{\rm fb} \simeq 1.0 M_{\odot} (M_{BH} / 3M_{\odot})^{-2/3}$.

For the golden sample of GRB-supernova association (GRBs 060218, 100316D, 091127, 120422A) in the {\it Swift} era (Zhang et al. 2012): three of them (except for GRB 120422A) lack X-ray plateaus. If some plateaus are the consequence of the fall-back accretion, then it can well explain the lacking plateaus with bright SN since the fall-back material cannot survive during the explosion. In our model, GRB 081029 and GRB 100621A might be accompanied with a low energy supernova, or even a failed supernova.

The later light curves, especially those with unusual features, can indeed help to learn about the central engines. In the BH accretion system, the fall-back process seems plausible under some conditions. Based on this scenario (or other possible scenarios that can provide such a delayed energy), we propose the delayed energy injection model to interpret the later bump in the multiple-band observation of afterglows. However, we have only considered the simplified case. Some physical processes for photons streaming through the external shock may be considered later. Besides, the jet power from the fall-back accretion should depend on the spin evolution of the BH and the magnetic field around the BH. Our model can be tested using more afterglow samples with observations in more bands and detailed spectral evolution data.

\acknowledgments

\appendix
We thank Shanqin Wang, Shujing Hou, Kai Wang and Yuanpei Yang for helpful discussion. We also thank Nardini et al. and Greiner at al. for kindly sharing the data of GRB 081029 and GRB 100621A obtained with the seven-channel Gamma-Ray burst Optical and Near-infrared Detector (GROND). Our study made use of data supplied by the UK Swift Science Data Centre at the University of Leicester. This work was supported by the National Basic Research Program of China (973 Program, Grant No. 2009CB824800 and No. 2013CB834900) and the National Natural Science Foundation of China (Grant No. 11033002). X. F. Wu acknowledges support by the One-Hundred-
Talents Program and the Youth Innovation Promotion Association of Chinese Academy of Sciences.

\clearpage

\begin{deluxetable}{ccccc}
\tabletypesize{\scriptsize}
\tablewidth{0pt}
\tablecaption{Beat fit parameters for the multiple-wavelength light curves of GRB 081029.\label{TABLE:Fit1}}

\tablehead{%
        \colhead{Fireball parameter} &
        \colhead{Value} &
        \colhead{Injection Parameter} &
        \colhead{broken-power-law mode} &
        \colhead{``top-hat'' mode}}
\startdata
$\theta_j$ (rad) & 0.03    &  $L_{\rm p}$ (erg s$^{-1}$) & $4.5 \times 10^{50}$  &  \\
$\Gamma_0$       & 400     &  $L_0$ (erg s$^{-1}$)     &         &       $4.0 \times 10^{50}$       \\
$E_{K,\mathrm{iso}}$ (erg) & $1.2 \times 10^{52}$  &  $t_{\rm s}$ (s)  &   3300.0   &  10000.0      \\
$p$              & 2.4     &  $t_{\rm p}$ (s)      &  3700.0    &   \\
$\epsilon_{e}$             & 0.028   & $t_{\rm e}$ (s)   &       3900.0       &    11000.0 \\
$\epsilon_{B}$             & 0.5     & $\alpha_{r}$  &  0.5  &   \\
$n$ (cm$^{-3}$)            & 10.0    & $\alpha_{d}$  &  -1.3 &   \\
$z$                      & 3.8479    &   $s$       &   0.5 &     \\
\enddata

\tablecomments{We use the injection of broken-power-law mode to interpret the giant bump. The second injection of ``top-hat'' mode corresponds to the flat tail. The time parameters are all defined in the observer frame, whereas the isotropic luminosity parameters ($L_{\rm p}$, $L_0$) are defined in the cosmological local frame.}
\end{deluxetable}

\begin{deluxetable}{ccccc}
\tabletypesize{\scriptsize}
\tablewidth{0pt}
\tablecaption{Beat fit parameters for the multiple-wavelength light curves of GRB 100621A.\label{TABLE:Fit2}}

\tablehead{%
        \colhead{Fireball parameter} &
        \colhead{off-axis jet} &
        \colhead{on-axis jet} &
        \colhead{Injection Parameter} &
        \colhead{broken-power-law mode}}
\startdata
$\theta_{obs}$\tablenotemark{\dag} (rad) &  0.24   &  0.0   & $L_{\rm p}$ (erg s$^{-1}$) & $5.0 \times 10^{49}$  \\
$\theta_j$ (rad)     & 0.2     &  0.04  & $t_{\rm s}$ (s)  &   3600.0    \\
$\Gamma_0$           & 100     &  70    & $t_{\rm p}$ (s)  &   4100.0    \\
$E_{K,\mathrm{iso}}$ (erg) & $5.0 \times 10^{52}$  & $1.0 \times 10^{52}$ & $t_{\rm e}$ (s)&  4600.0 \\
$p$                  & 2.1     &  2.2   & $\alpha_{r}$      &  0.5       \\
$\epsilon_{e}$       & 0.04    & 0.04   & $\alpha_{d}$      &  -1.2      \\
$\epsilon_{B}$       & 0.1     & 0.06   &    $s$            &  0.5       \\
$n$ (cm$^{-3}$)      & 2.0     &  2.0   &                   &            \\
$z$                  & 0.542   & 0.542  &                   &            \\
\enddata

\tablenotetext{\dag}{$\theta_{obs}$ is defined as the angle between the line of
sight and the jet axis.}

\tablecomments{We use the on-axis jet with injection of broken-power-law mode to interpret the intensive jump at $\sim 4.5$ ks, while the off-axis jet will explain the slow decreasing flux at the later time. The time parameters are all defined in the observer frame, whereas the isotropic luminosity parameters ($L_{\rm p}$) are defined in the cosmological local frame.}
\end{deluxetable}

\begin{deluxetable}{cccccccc}
\tabletypesize{\scriptsize}
\tablewidth{0pt}
\tablecaption{The value of dust extinction of seven bands used in the fitting of GRB 100621A.\label{TABLE:Ext}}

\tablehead{%
        \colhead{} &
        \colhead{$g'$} &
        \colhead{$r'$} &
        \colhead{$i'$} &
        \colhead{$z'$} &
        \colhead{$J$} &
        \colhead{$H$} &
        \colhead{$K$}}
\startdata
$A$ (mag) & 6.51 & 5.20 & 4.13 & 3.31 & 1.95 & 1.19 & 0.86 \\
\enddata
\end{deluxetable}

\begin{figure}
   \plottwo{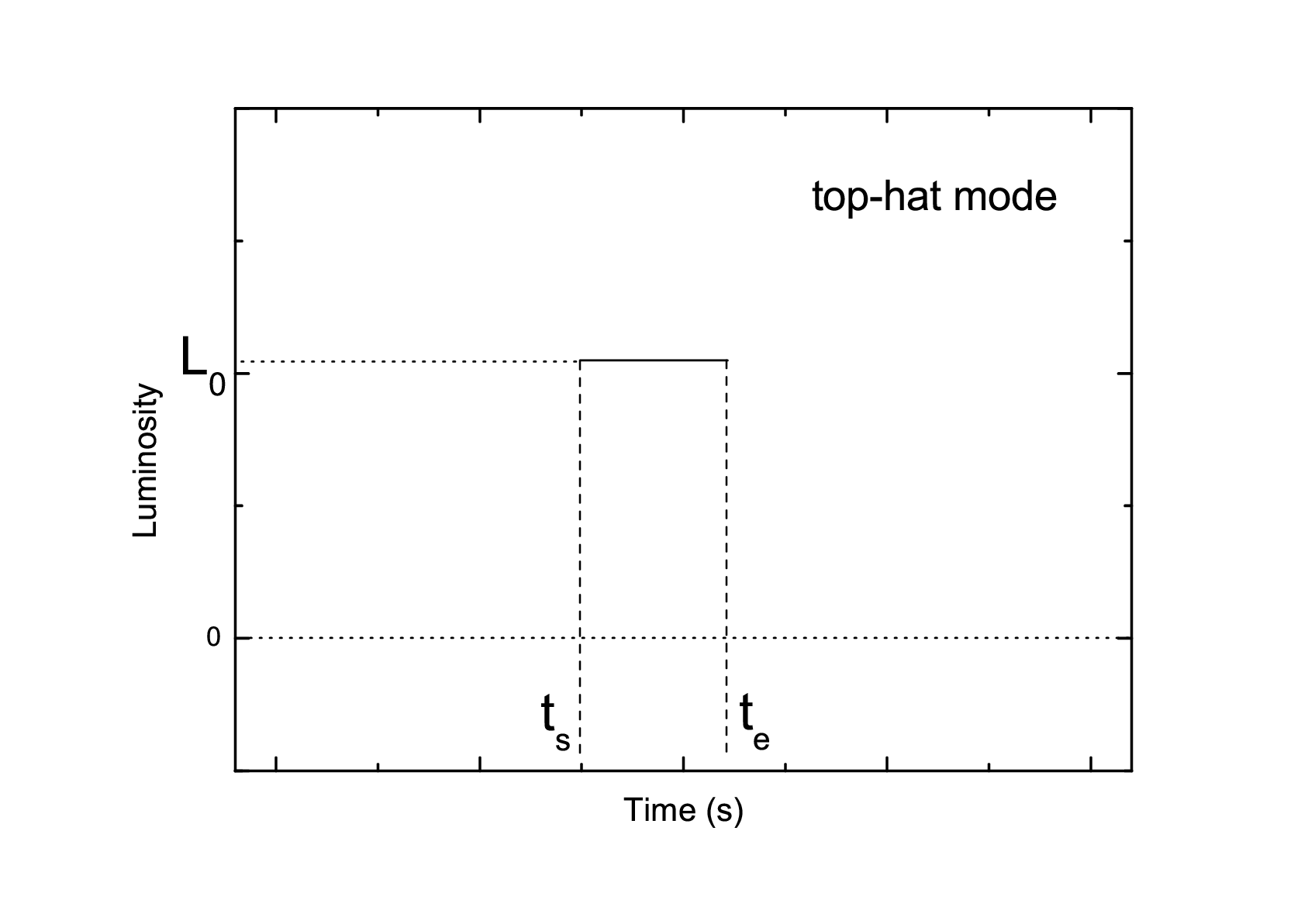}{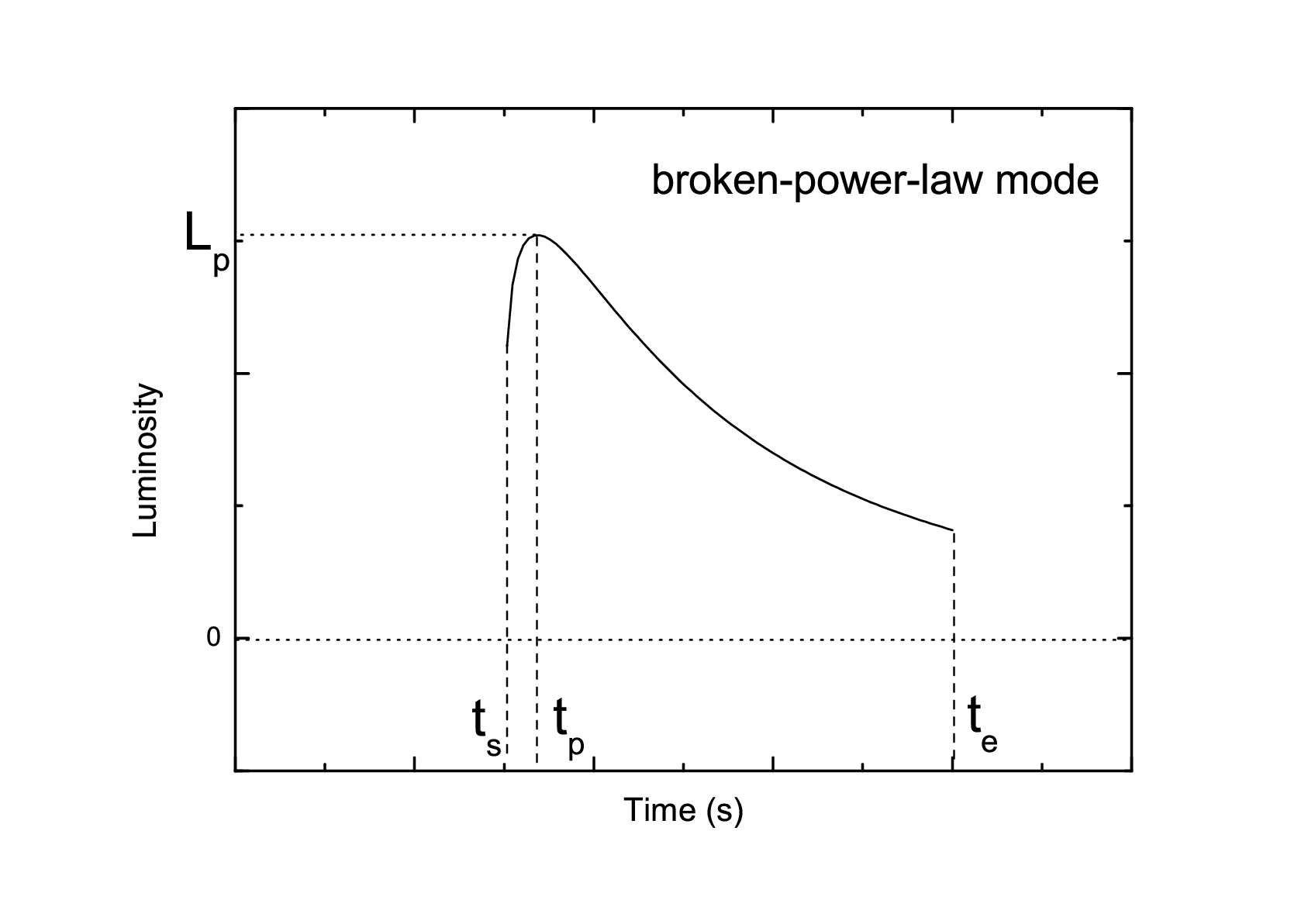}
   \caption{Left panel: A schematic sketch of the ``top-hat'' injection mode. $L_0$ is the constant injection luminosity. $t_{\rm s}$ and $t_{\rm e}$ refer to the start time and end time of the injection period respectively.
Right panel: A schematic sketch of the broken-power-law injection mode. According to Equation (4), the injection starts from $t_{\rm s}$ and rises to the peak luminosity $L_{\rm p}$ at $t_{\rm p}$. The injection ends at some time $t_{\rm e}$.}
   \label{Fig:plot1}
\end{figure}

\begin{figure}
   \begin{center}
   \plotone{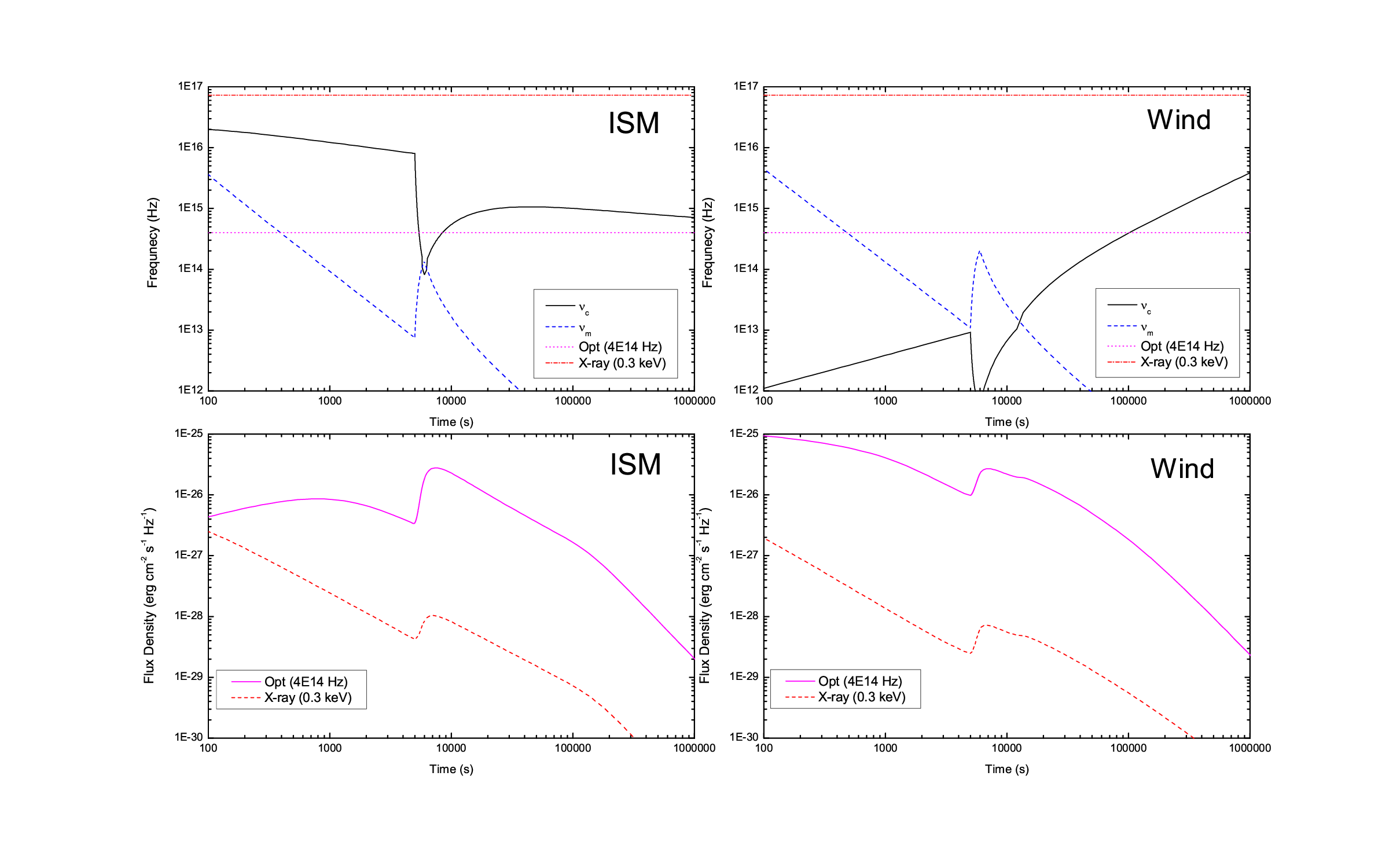}
   \caption{Results in the ``top-hat'' mode. The top left panel shows the evolution of two frequencies, $\nu_m$ and $\nu_c$ in the ISM case. The two straight lines note the two typical bands ($4.0 \times 10^{14}$ Hz as optical band and $0.3$ keV as soft X-ray band) in which the afterglows are calculated. The lower left panel shows the afterglows in these two bands (redshift $z = 1$ is assumed). The top right panel and lower right panel show similar plots as to left panels, but the environment is wind type.}
   \label{Fig:plot2}
   \end{center}
\end{figure}

\begin{figure}
   \begin{center}
   \plotone{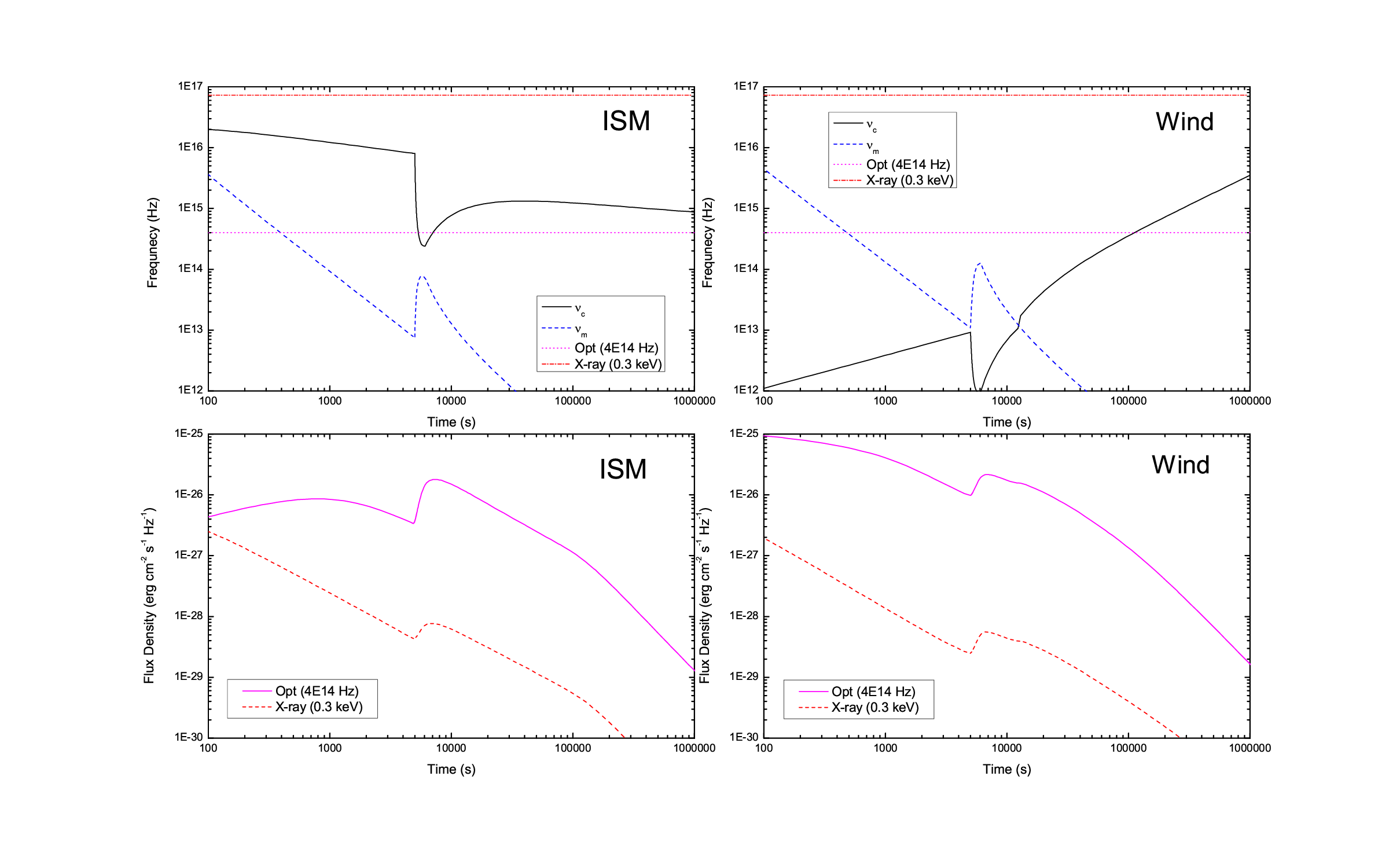}
   \caption{Results in the broken-power-law mode. The top left panel shows the evolution of two frequencies, $\nu_m$ (dashed line) and $\nu_c$ (solid line) in the ISM case. The two straight lines note the two typical bands ($4.0 \times 10^{14}$ Hz as optical band and $0.3$ keV as soft X-ray band) in which the afterglows are calculated. The lower left panel shows the afterglows in these two bands (redshift $z = 1$ is assumed). The top right panel and lower right panel show the similar plots as left panels, but the environment is wind type.}
   \label{Fig:plot3}
   \end{center}
\end{figure}

\begin{figure}
   \begin{center}
   \includegraphics[scale=0.3]{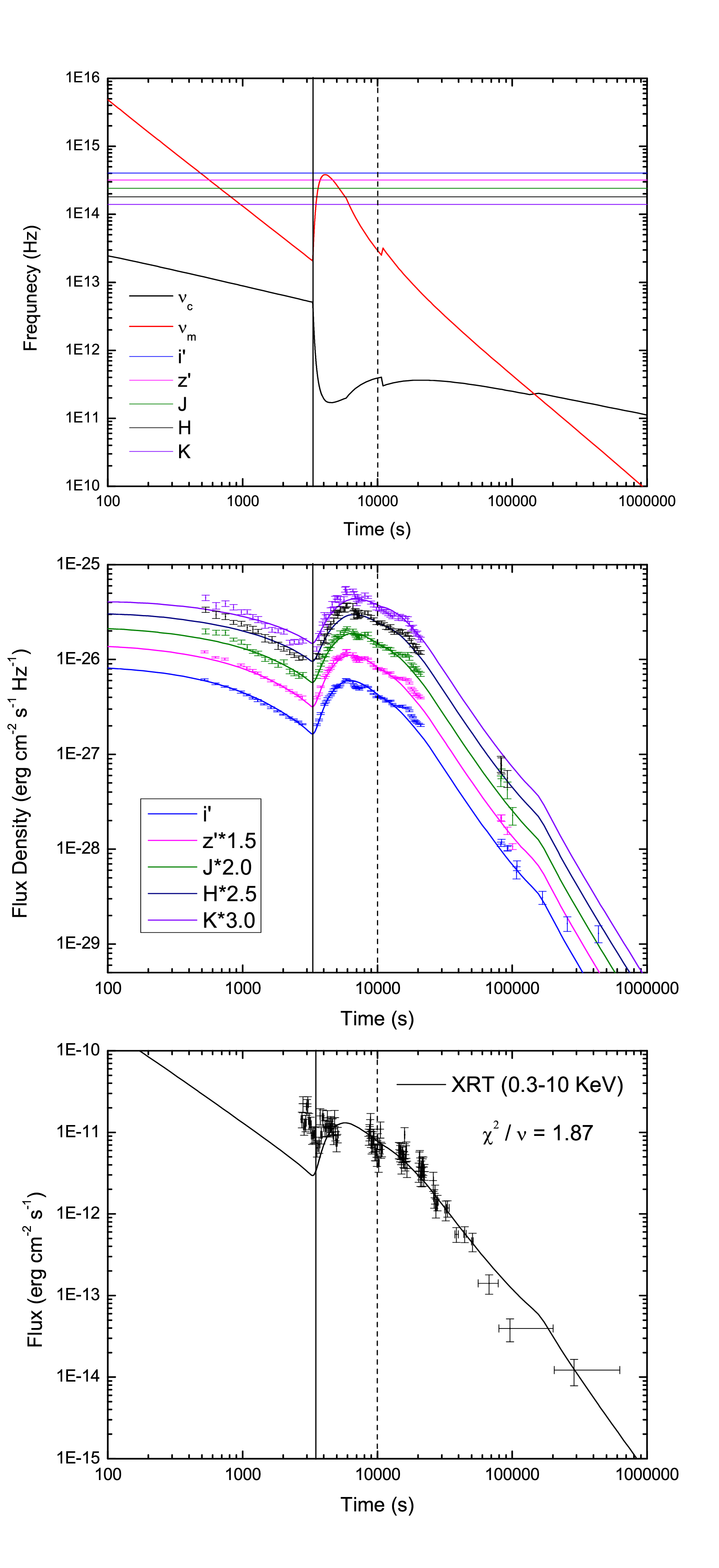}
   \caption{Our best fit to the multi-band afterglow of GRB 081029, by assuming two energy injections. The top panel shows the evolution of $\nu_m$ (dashed line) and $\nu_c$ (solid line), the middle panel shows the fitting result in five optical and infrared bands (the value of four lower bands have been multiplied by four different factors to display better), the lower panel shows the fitting result in XRT band. The vertical solid line notes the starting time of the first injection, and the dashed line notes the starting time of the second injection. Detailed parameters are listed in Tab. 1. Data are taken from Nardini et al. (2011a).}
   \label{Fig:plot4}
   \end{center}
\end{figure}

\begin{figure}
   \begin{center}
   \includegraphics[scale=0.3]{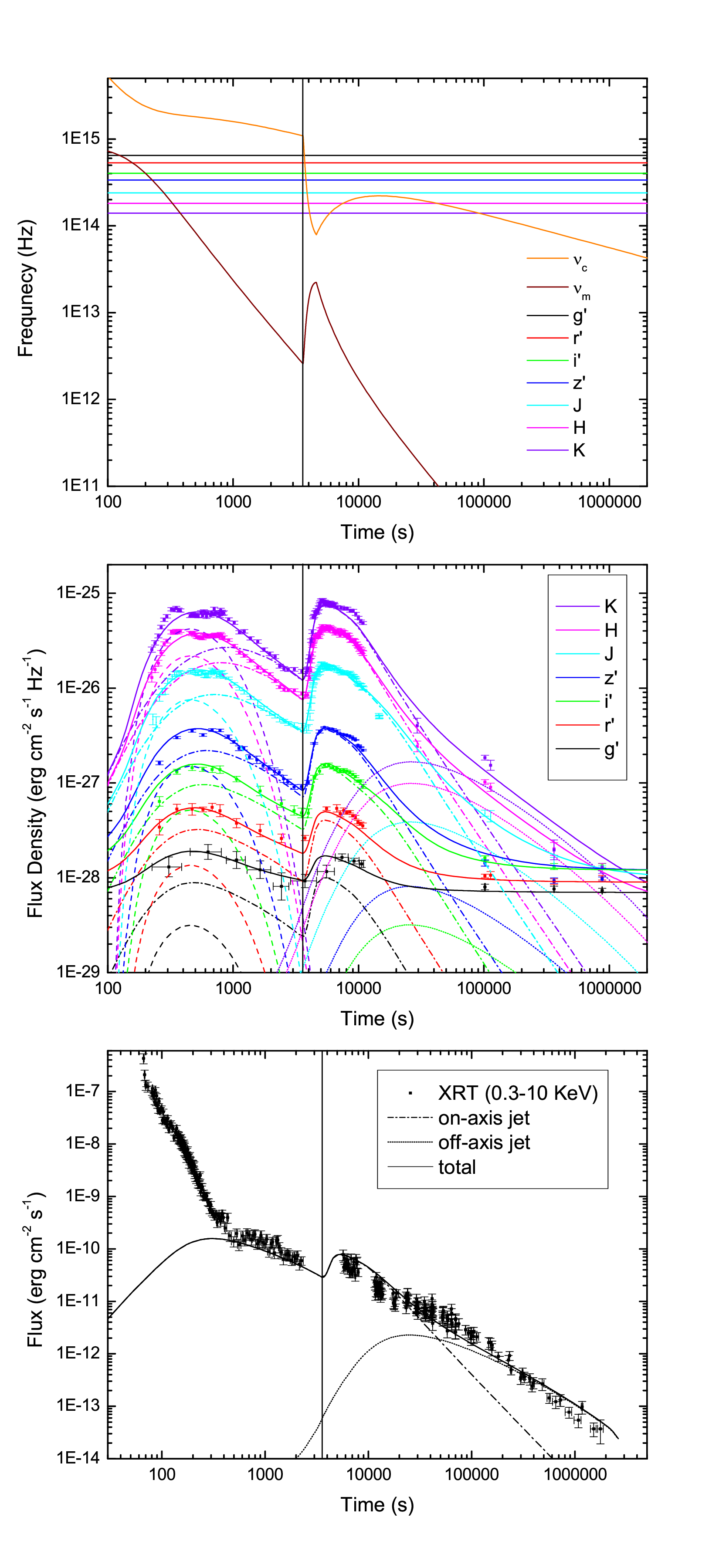}
   \caption{Our best fit to the multi-band afterglow of GRB 100621A, using three components (flare, on-axis jet with delayed injection and off-axis jet). The top panel shows the evolution of $\nu_m$ and $\nu_c$ of the on-axis jet. The middle panel shows the fitting result of seven bands of GROND, the dashed lines mark the flare, the dash-dotted lines mark the afterglow of on-axis jet, the short-dotted lines mark the afterglow of off-axis jet and the solid lines show the total flux of these three components. The lower panel shows the fitting result in XRT band. The vertical solid line notes the starting time of the injection. Detailed parameters are listed in Tab. 2, 3. GROND data are taken from Greiner et al. (2013).}
   \label{Fig:plot5}
   \end{center}
\end{figure}


\clearpage

\begin{thebibliography}{99}
\bibitem[Bernardini et
al.(2012)]{2012A&A...539A...3B} Bernardini, M.~G., Margutti, R., Mao, J., Zaninoni, E., \& Chincarini, G.\ 2012, \aap, 539, A3

\bibitem[Blandford
\& Payne(1982)]{1982MNRAS.199..883B} Blandford, R.~D., \& Payne, D.~G.\ 1982, \mnras, 199, 883

\bibitem[Blandford
\& Znajek(1977)]{1977MNRAS.179..433B} Blandford, R.~D., \& Znajek, R.~L.\ 1977, \mnras, 179, 433

\bibitem[Burrows et al.(2005)]{2005SSRv..120..165B} Burrows, D.~N., Hill,
J.~E., Nousek, J.~A., et al.\ 2005, \ssr, 120, 165

\bibitem[Cannizzo et al.(2011)]{2011ApJ...734...35C} Cannizzo, J.~K.,
Troja, E., \& Gehrels, N.\ 2011, \apj, 734, 35

\bibitem[Dai(2004)]{2004ApJ...606.1000D} Dai, Z.~G.\ 2004, \apj, 606, 1000

\bibitem[Dai
\& Liu(2012)]{2012ApJ...759...58D} Dai, Z.~G., \& Liu, R.-Y.\ 2012, \apj, 759, 58

\bibitem[Dai
\& Lu(1998)]{1998A&A...333L..87D} Dai, Z.~G., \& Lu, T.\ 1998a, \aap, 333, L87

\bibitem[Dai
\& Lu(1998)]{1998PhRvL..81.4301D} Dai, Z.~G., \& Lu, T.\ 1998b, Physical Review Letters, 81, 4301

\bibitem[Dai
\& Wu(2003)]{2003ApJ...591L..21D} Dai, Z.~G., \& Wu, X.~F.\ 2003, \apjl, 591, L21

\bibitem[D'Elia et al.(2008)]{2008GCN..8438....1D} D'Elia, V., Covino, S.,
\& D'Avanzo, P.\ 2008, GRB Coordinates Network, 8438, 1

\bibitem[Fan
\& Xu(2006)]{2006MNRAS.372L..19F} Fan, Y.-Z., \& Xu, D.\ 2006, \mnras, 372, L19

\bibitem[Filgas et
al.(2012)]{2012A&A...546A.101F} Filgas, R., Greiner, J., Schady, P., et al.\ 2012, \aap, 546, A101

\bibitem[Freedman
\& Waxman(2001)]{2001ApJ...547..922F} Freedman, D.~L., \& Waxman, E.\ 2001, \apj, 547, 922

\bibitem[Gehrels et al.(2004)]{2004ApJ...611.1005G} Gehrels, N.,
Chincarini, G., Giommi, P., et al.\ 2004, \apj, 611, 1005

\bibitem[Gehrels et al.(2009)]{2009ARA&A..47..567G} Gehrels, N.,
Ramirez-Ruiz, E., \& Fox, D.~B.\ 2009, \araa, 47, 567

\bibitem[Granot et al.(1999)]{1999ApJ...513..679G} Granot, J., Piran, T.,
\& Sari, R.\ 1999, \apj, 513, 679

\bibitem[Greiner et al.(2013)]{2013arXiv1304.5852G} Greiner, J.,
Kr{\"u}hler, T., Nardini, M., et al.\ 2013, arXiv:1304.5852

\bibitem[He et al.(2009)]{2009ApJ...706.1152H} He, H.-N., Wang, X.-Y., Yu,
Y.-W., \& M{\'e}sz{\'a}ros, P.\ 2009, \apj, 706, 1152

\bibitem[Holland et al.(2012)]{2012ApJ...745...41H} Holland, S.~T., De
Pasquale, M., Mao, J., et al.\ 2012, \apj, 745, 41

\bibitem[Huang et al.(1999)]{1999MNRAS.309..513H} Huang, Y.~F., Dai, Z.~G.,
\& Lu, T.\ 1999, \mnras, 309, 513

\bibitem[Huang et al.(2000)]{2000MNRAS.316..943H} Huang, Y.~F., Dai, Z.~G.,
\& Lu, T.\ 2000a, \mnras, 316, 943

\bibitem[Huang et al.(2000)]{2000ApJ...543...90H} Huang, Y.~F., Gou, L.~J.,
Dai, Z.~G., \& Lu, T.\ 2000b, \apj, 543, 90

\bibitem[Huang et al.(2007)]{2007ChJAA...7..397H} Huang, Y.-F., Lu, Y.,
Wong, A.~Y.~L., \& Cheng, K.~S.\ 2007, \cjaa, 7, 397

\bibitem[Kong
\& Huang(2010)]{2010ScChG..53S..94K} Kong, S., \& Huang, Y.\ 2010, Science in China G: Physics and Astronomy, 53, 94

\bibitem[Kong et al.(2010)]{2010MNRAS.402..409K} Kong, S.~W., Wong,
A.~Y.~L., Huang, Y.~F., \& Cheng, K.~S.\ 2010, \mnras, 402, 409

\bibitem[Kumar et al.(2013)]{2013arXiv1304.1545K} Kumar, P., Barniol Duran,
R., Bosnjak, Z., \& Piran, T.\ 2013, arXiv:1304.1545

\bibitem[Kumar et al.(2008)]{2008Sci...321..376K} Kumar, P., Narayan, R.,
\& Johnson, J.~L.\ 2008a, Science, 321, 376

\bibitem[Kumar et al.(2008)]{2008MNRAS.388.1729K} Kumar, P., Narayan, R.,
\& Johnson, J.~L.\ 2008b, \mnras, 388, 1729

\bibitem[Lazzati
\& Perna(2007)]{2007MNRAS.375L..46L} Lazzati, D., \& Perna, R.\ 2007, \mnras, 375, L46

\bibitem[Lazzati et
al.(2002)]{2002A&A...396L...5L} Lazzati, D., Rossi, E., Covino, S., Ghisellini, G., \& Malesani, D.\ 2002, \aap, 396, L5

\bibitem[Lee et al.(2000)]{2000PhR...325...83L} Lee, H.~K., Wijers,
R.~A.~M.~J., \& Brown, G.~E.\ 2000, \physrep, 325, 83

\bibitem[Lei et al.(2009)]{2009ApJ...700.1970L} Lei, W.~H., Wang, D.~X.,
Zhang, L., et al.\ 2009, \apj, 700, 1970

\bibitem[Lei et al.(2008)]{2008ChJAA...8..404L} Lei, W.-H., Wang, D.-X.,
Zou, Y.-C., \& Zhang, L.\ 2008, \cjaa, 8, 404

\bibitem[Lei et al.(2013)]{2013ApJ...765..125L} Lei, W.-H., Zhang, B.,
\& Liang, E.-W.\ 2013, \apj, 765, 125

\bibitem[Lindner et al.(2010)]{2010ApJ...713..800L} Lindner, C.~C.,
Milosavljevi{\'c}, M., Couch, S.~M., \& Kumar, P.\ 2010, \apj, 713, 800

\bibitem[Lindner et al.(2012)]{2012ApJ...750..163L} Lindner, C.~C.,
Milosavljevi{\'c}, M., Shen, R., \& Kumar, P.\ 2012, \apj, 750, 163

\bibitem[MacFadyen
\& Woosley(1999)]{1999ApJ...524..262M} MacFadyen, A.~I., \& Woosley, S.~E.\ 1999, \apj, 524, 262

\bibitem[MacFadyen et al.(2001)]{2001ApJ...550..410M} MacFadyen, A.~I.,
Woosley, S.~E., \& Heger, A.\ 2001, \apj, 550, 410

\bibitem[Mao et
al.(2010)]{2010A&A...518A..27M} Mao, Z., Yu, Y.~W., Dai, Z.~G., Pi, C.~M., \& Zheng, X.~P.\ 2010, \aap, 518, A27

\bibitem[McKinney(2005)]{2005ApJ...630L...5M} McKinney, J.~C.\ 2005, \apjl,
630, L5

\bibitem[Metzger et al.(2011)]{2011MNRAS.413.2031M} Metzger, B.~D.,
Giannios, D., Thompson, T.~A., Bucciantini, N.,
\& Quataert, E.\ 2011, \mnras, 413, 2031

\bibitem[Milvang-Jensen et al.(2010)]{2010GCN..10876...1M} Milvang-Jensen,
B., Goldoni, P., Tanvir, N.~R., et al.\ 2010, GRB Coordinates Network,
10876, 1

\bibitem[Nakar
\& Granot(2007)]{2007MNRAS.380.1744N} Nakar, E., \& Granot, J.\ 2007, \mnras, 380, 1744

\bibitem[Nakar
\& Piran(2003)]{2003ApJ...598..400N} Nakar, E., \& Piran, T.\ 2003, \apj, 598, 400

\bibitem[Narayan et al.(2013)]{2013arXiv1303.3004N} Narayan, R., McClintock, J.~E., \& Tchekhovskoy, A.\ 2013, arXiv:1303.3004

\bibitem[Narayan et al.(2001)]{2001ApJ...557..949N} Narayan, R., Piran, T.,
\& Kumar, P.\ 2001, \apj, 557, 949

\bibitem[Nardini et al.(2011)]{2011AIPC.1358..150N} Nardini, M., Greiner,
J., Klose, S., et al.\ 2011b, American Institute of Physics Conference
Series, 1358, 150

\bibitem[Nardini et
al.(2011)]{2011A&A...531A..39N} Nardini, M., Greiner, J., Kr{\"u}hler, T., et al.\ 2011a, \aap, 531, A39

\bibitem[Nava et al.(2013)]{2013MNRAS.tmp.1563N} Nava, L., Sironi, L.,
Ghisellini, G., Celotti, A., \& Ghirlanda, G.\ 2013, \mnras, 1563

\bibitem[Norris et al.(2005)]{2005ApJ...627..324N} Norris, J.~P., Bonnell,
J.~T., Kazanas, D., et al.\ 2005, \apj, 627, 324

\bibitem[Pe'er(2012)]{2012ApJ...752L...8P} Pe'er, A.\ 2012, \apjl, 752, L8

\bibitem[Perna et al.(2006)]{2006ApJ...636L..29P} Perna, R., Armitage,
P.~J., \& Zhang, B.\ 2006, \apjl, 636, L29

\bibitem[Piran et al.(1993)]{1993MNRAS.263..861P} Piran, T., Shemi, A.,
\& Narayan, R.\ 1993, \mnras, 263, 861

\bibitem[Popham et al.(1999)]{1999ApJ...518..356P} Popham, R., Woosley,
S.~E., \& Fryer, C.\ 1999, \apj, 518, 356

\bibitem[Proga
\& Zhang(2006)]{2006MNRAS.370L..61P} Proga, D., \& Zhang, B.\ 2006, \mnras, 370, L61

\bibitem[Sari
\& Esin(2001)]{2001ApJ...548..787S} Sari, R., \& Esin, A.~A.\ 2001, \apj, 548, 787

\bibitem[Sari
\& Piran(1999)]{1999ApJ...517L.109S} Sari, R., \& Piran, T.\ 1999a, \apjl, 517, L109

\bibitem[Sari
\& Piran(1999)]{1999ApJ...520..641S} Sari, R., \& Piran, T.\ 1999b, \apj, 520, 641

\bibitem[Sari et al.(1998)]{1998ApJ...497L..17S} Sari, R., Piran, T.,
\& Narayan, R.\ 1998, \apjl, 497, L17

\bibitem[Tutukov
\& Fedorova(2007)]{2007ARep...51..291T} Tutukov, A.~V., \& Fedorova, A.~V.\ 2007, Astronomy Reports, 51, 291

\bibitem[Ukwatta et al.(2010)]{2010GCNR..291....1U} Ukwatta, T.~N.,
Stratta, G., Evans, P.~A., et al.\ 2010, GCN Report, 291, 1

\bibitem[Wang \& Cheng(2012)]{2012MNRAS.421..908W} Wang, F.~Y., \& Cheng, K.~S.\ 2012, \mnras, 421, 908

\bibitem[Waxman(1997)]{1997ApJ...491L..19W} Waxman, E.\ 1997, \apjl, 491,
L19

\bibitem[Woosley(1993)]{1993ApJ...405..273W} Woosley, S.~E.\ 1993, \apj,
405, 273

\bibitem[Wu et al.(2003)]{2003MNRAS.342.1131W} Wu, X.~F., Dai, Z.~G.,
Huang, Y.~F., \& Lu, T.\ 2003, \mnras, 342, 1131

\bibitem[Wu et al.(2013)]{2013ApJ...767L..36W} Wu, X.-F., Hou, S.-J.,
\& Lei, W.-H.\ 2013, \apjl, 767, L36

\bibitem[Xie et al.(2009)]{2009MNRAS.398..583X} Xie, Y., Huang, Z.-Y., Jia,
X.-F., Fan, S.-J., \& Liu, F.-F.\ 2009, \mnras, 398, 583

\bibitem[Xu \& Huang(2010)]{2010A&A...523A...5X} Xu, M., \& Huang, Y.~F.\ 2010, \aap, 523, A5

\bibitem[Xue et al.(2009)]{2009A&A...498..671X} Xue, R.-R., Fan, Y.-Z., \& Wei, D.-M.\ 2009, \aap, 498, 671

\bibitem[Xue-Wen et al.(2009)]{2009arXiv0907.1767X} Xue-Wen, L., Xue-Feng,
W., \& Tan, L.\ 2009, arXiv:0907.1767

\bibitem[Yu
\& Huang(2013)]{2013arXiv1303.2161Y} Yu, Y.~B., \& Huang, Y.~F.\ 2013, arXiv:1303.2161

\bibitem[Yu
\& Dai(2007)]{2007A&A...470..119Y} Yu, Y.~W., \& Dai, Z.~G.\ 2007, \aap, 470, 119

\bibitem[Yuan
\& Zhang(2012)]{2012ApJ...757...56Y} Yuan, F., \& Zhang, B.\ 2012, \apj, 757, 56

\bibitem[Zhang(2007)]{2007ChJAA...7....1Z} Zhang, B.\ 2007, \cjaa, 7, 1

\bibitem[Zhang et al.(2006)]{2006ApJ...642..354Z} Zhang, B., Fan, Y.~Z.,
Dyks, J., et al.\ 2006, \apj, 642, 354

\bibitem[Zhang
\& M{\'e}sz{\'a}ros(2001)]{2001ApJ...552L..35Z} Zhang, B., \& M{\'e}sz{\'a}ros, P.\ 2001, \apjl, 552, L35

\bibitem[Zhang et al.(2012)]{2012ApJ...756..190Z} Zhang, B.-B., Fan, Y.-Z.,
Shen, R.-F., et al.\ 2012, \apj, 756, 190

\bibitem[Zhang et al.(2008)]{2008ApJ...679..639Z} Zhang, W., Woosley,
S.~E., \& Heger, A.\ 2008, \apj, 679, 639

\bibitem[Zou et al.(2013)]{2013arXiv1307.2650Z} Zou, Y.~C., Wang, F.~Y.,
\& Cheng, K.~S.\ 2013, arXiv:1307.2650
\end{thebibliography}
\end{document}